\documentclass[%
reprint,
superscriptaddress,.
amsmath,
amssymb,
aps,
pra,
longbibliography,
floatfix,
]{revtex4-2}

\usepackage{mathrsfs,braket}
\usepackage{amssymb, amsbsy, amsmath, latexsym, dsfont, array, layout, graphicx,mathrsfs,color, bm,mathtools}
\usepackage{physics}
\usepackage[dvipsnames]{xcolor}
\usepackage[unicode=true,breaklinks=true,colorlinks=true]{hyperref}
\usepackage{bigints}
\usepackage[T1]{fontenc}
\usepackage{newtxtext}
\usepackage[utf8]{inputenc}

\hypersetup{citecolor=blue,urlcolor=blue,linkcolor=blue,filecolor=blue}
\newcommand{\fig}[1]{Fig.~\ref{#1}}
\newcommand{\figg}[1]{Figure~\ref{#1}}

\usepackage[normalem]{ulem}

\begin{document}

\title{Strong coherent ion-electron coupling using a wire data bus}

\author{Baiyi Yu}
\author{Ralf Betzholz}
\author{Jianming Cai}
\email{jianmingcai@hust.edu.cn}
\affiliation{School of Physics, Hubei Key Laboratory of Gravitation and Quantum Physics, International Joint Laboratory on Quantum Sensing and Quantum Metrology, Institute for Quantum Science and Engineering, Huazhong University of Science and Technology, Wuhan 430074, China}

\begin{abstract}
Ion-ion coupling over long distances represents a highly useful resource for quantum technologies, for example, to sympathetically cool or interconnect qubits in ion-based quantum-computing architectures.
In this respect, the recently demonstrated wire-mediated ion-ion coupling stands due to the simplification of its trap layout and its prospects for deterministic entanglement.
However, the strength of such coherent ion-wire-ion coupling is typically weak, hindering its practical utilization.
Here, we propose a wire-mediated scheme for coherent ion-electron coupling. The scheme not only enables the sympathetic cooling of electrons via advanced ion-cooling techniques, but also allows to promote the effective ion-ion coupling strength by orders of magnitudes via electron mediation. 
Our work thus paves a way toward quantum information processing in ion-electron hybrid quantum systems.
\end{abstract}

\date{July 29, 2024}

\maketitle
\section{Introduction}
Coherent coupling represents a paramount ingredient for the implementation of most quantum technologies~\cite{gaebelRoomtemperatureCoherentCoupling2006,brownCoupledQuantizedMechanical2011,harlanderTrappedionAntennaeTransmission2011,caiLargescaleQuantumSimulator2013,wilsonTunableSpinSpin2014,viennotCoherentCouplingSingle2015}, because it is crucial for establishing one of their most essential resources, namely the entanglement of different quantum-mechanical degrees of freedom. For example, the coupling between qubits is an integral ingredient for the realization of entangling gates in quantum computing~\cite{ciracQuantumComputationsCold1995,monroeDemonstrationFundamentalQuantum1995,poyatosQuantumGatesHot1998,molmerMultiparticleEntanglementHot1999,sorensenQuantumComputationIons1999,sackettExperimentalEntanglementFour2000a,schmidt-kalerRealizationCiracZoller2003,roosControlMeasurementThreeQubit2004}. On the other hand, besides the coupling of different degrees of freedom within the same system, the coherent coupling of entirely different types of quantum systems is likewise intriguing, since it opens routes to build hybrid platforms, potentially harnessing and combining the advantages of different quantum systems~\cite{schmidtSpectroscopyUsingQuantum2005,zhuCoherentCouplingSuperconducting2011,stuteTunableIonPhoton2012,daniilidisQuantumInformationProcessing2013,daniilidisQuantumInterfacesAtomic2013,caiHybridSensorsBased2014,tabuchiCoherentCouplingFerromagnetic2015,ballanceHybridQuantumLogic2015,tanMultielementLogicGates2015,wangSinglequbitQuantumMemory2017,kotlerHybridQuantumSystems2017,footTwofrequencyOperationPaul2018,kargLightmediatedStrongCoupling2020a,weckesserObservationFeshbachResonances2021}.

Trapped-ion systems, with their rich internal energy levels, for instance, were shown to exhibit unique advantages in quantum information processing~\cite{ciracScalableQuantumComputer2000,guldeImplementationDeutschJozsa2003,haffnerScalableMultiparticleEntanglement2005,benhelmFaulttolerantQuantumComputing2008,kimQuantumSimulationFrustrated2010,lanyonUniversalDigitalQuantum2011,blattQuantumSimulationsTrapped2012,brittonEngineeredTwodimensionalIsing2012,kokailSelfverifyingVariationalQuantum2019,pinoDemonstrationTrappedionQuantum2021,ringbauerUniversalQuditQuantum2022,chenScalableProgrammablePhononic2023,katzDemonstrationThreeFourbody2023,qiaoTunableQuantumSimulation2024} 
and precision sensing~\cite{burdQuantumAmplificationMechanical2019,mccormickQuantumenhancedSensingSingleion2019,gilmoreQuantumenhancedSensingDisplacements2021a,carneyTrappedElectronsIons2021a,budkerMillichargedDarkMatter2022,sailerMeasurementBoundelectronGfactor2022}, 
whereas trapped-electron systems, with their simpler spin-up and -down internal states, have a long history in high-precision measurements~\cite{brownGeoniumTheoryPhysics1986,vandyckNewHighprecisionComparison1987,sturmHighprecisionMeasurementAtomic2014,cridlandSingleMicrowavePhoton2016,farnhamDeterminationElectronAtomic1995,peilObservingQuantumLimit1999,mittlemanBoundMathitCPTLorentz1999,odomNewMeasurementElectron2006,hannekeNewMeasurementElectron2008}
and hold the potential to accelerate quantum information processing speed due to their remarkably large charge-to-mass ratio, as compared to ions~\cite{platzmanQuantumComputingElectrons1999,ciaramicoliScalableQuantumProcessor2003,zurita-sanchezWiringSingleElectron2008,marzoliExperimentalTheoreticalChallenges2009,schusterProposalManipulatingDetecting2010,yangCouplingEnsembleElectrons2016,matthiesenTrappingElectronsRoomTemperature2021,osadaFeasibilityStudyGroundstate2022,sutherlandOneTwoqubitGate2022a,yuFeasibilityStudyQuantum2022,zhouSingleElectronsSolid2022,kawakamiBlueprintQuantumComputing2023a,zhouElectronChargeQubit2023,yuEngineeringArtificialAtomic2024}. 
Therefore, coherently coupling trapped ions and electrons could not only make established techniques based on lasers, such as laser cooling~\cite{winelandLaserCoolingAtoms1979}, indirectly applicable to trapped electrons, 
but could also substantially enhance the ion-ion coupling strength if their interaction is mediated by electrons.

However, this task is not straightforwardly achievable due to the extreme mass difference and opposite electric charges of ions and electrons. 
These differences pose a great challenge for trapping electrons and ions in close proximity in order to exploit the direct Coulomb interaction between them.
Even in the scenario of a wire-mediated coupling~\cite{anCouplingTwoLasercooled2022}, where electrons and ions are separately trapped and connected via a conductor wire, the typically large detuning between their motional frequencies significantly suppresses their interaction.

In this work, we propose a scheme to establish such a wire-mediated coherent coupling between trapped ions and electrons that remedies their frequency detuning through a local parametric modulation of the electron trapping potential. Because of the remarkably greater charge-to-mass ratio of the electron, in feasible experimental setups, the  wire-mediated ion-electron coupling strength for single particles could be orders of magnitude larger than the wire-mediated ion-ion coupling strength~\cite{anCouplingTwoLasercooled2022}. 
Building on this basic hybridization, we further discuss an electron-mediated ion-ion coupling scheme, which not only promises a faster generation of entanglement, but would also allow sympathetic~\cite{bohmanSympatheticCoolingProtons2018,bohmanSympatheticCoolingTrapped2021,willSympatheticCoolingSchemes2022,willImagecurrentMediatedSympathetic2023} and even exchange cooling~\cite{heinzenQuantumlimitedCoolingDetection1990,fallekRapidExchangeCooling2024}. This, in turn, could promote precision measurements of charged particles, such as protons or antiprotons~\cite{tuTankCircuitAssistedCoupling2021,atrapcollaborationOneParticleMeasurementAntiproton2013,mooserDirectHighprecisionMeasurement2014,ulmerHighprecisionComparisonAntiprotontoproton2015,heisseHighPrecisionMeasurementProton2017,smorraPartsperbillionMeasurementAntiproton2017,niemannCryogenic9BeMathplus2019}.

\section{Ion-electron coupling scheme}\label{sec: ion-electron}
We consider a setup, schematically depicted in~\fig{fig1}(a), for the coherent coupling of ions and electrons, that are separately trapped at two distant positions and connected via a conductor wire (represented in golden).
The key problem that needs to be addressed is the large motional-frequency mismatch between the ions and the electrons for typical trapping potentials. This mismatch significantly suppresses the ion-electron coupling via the conductor wire.
To overcome this difficulty, we implement a proper electric driving field in the axial direction of the electron trap, as indicated by the potential modulation (gray) in \fig{fig1}(a).
This modulation results in an axial electron-motion component resonant with the axial ion motion, thereby effectively coupling the ions and electrons through a wire data bus despite the large mass difference between them.
For the case of each trap holding a single particle, the axial-motion Hamiltonian of the combined ion-wire-electron system can be written as 
\begin{equation}\label{eq:H}
    H=H_i+H_e+H_{I},
\end{equation}
where $H_{i(e)}$ is the Hamiltonian of the axial ion (electron) motion and $H_I$ describes the wire-mediated ion-electron coupling.
Although Eq.~\eqref{eq:H} describes only the axial motion, the results remain valid when considering the full equation of motion for both Paul and Penning traps~\cite{tanSynchronizationParametricallyPumped1991,burdQuantumAmplificationMechanical2019,wittemerPhononPairCreation2019,fanMeasurementElectronMagnetic2023}.

\subsection{Hamiltonian and equivalent circuit}
We then derive the explicit form of the Hamiltonian for the case of each trap holding a single particle following Ref.~\cite{daniilidisWiringTrappedIons2009} and references therein.
Firstly, we consider a fictitious situation where the conductor wire is at potential $U_w$ and carries a total charge $Q_w=C_w U_w$, while the ion and the electron in each trap carry no charge. The effective distance $D_{i(e)}$ of the ion (electron) to the wire can be defined as
\begin{equation}\label{eq: effective distance}
    D_{i(e)}=\frac{U_w}{E_{i(e)}},
\end{equation}
where $E_{i(e)}$ is the electric field generated by the wire at the equilibrium position of the ion (electron) along the axial direction~\cite{anCouplingTwoLasercooled2022,willSympatheticCoolingSchemes2022}.
Then, the potential generated by the conductor wire at the ion (electron) position is given by 
\begin{equation}\label{eq:eq:U ion electron}
    U_{i(e)}=U_{i(e),0}-x_{i(e)} \frac{U_w}{D_{i(e)}}, 
\end{equation}
where $x_{i(e)}$ is the position of the ion (electron) relative to its equilibrium point and $U_{i(e),0}$ is the potential generated by the conductor wire at the equilibrium point. One can then change to a dual case, where the wire has zero net charge and is at potential $U_w^\prime$, while the ion (electron) carries the charge $q_{i(e)}$.
Application of Green's reciprocity theorem to these two cases leads to
\begin{equation}\label{eq: U wire prime}
    U_w^\prime=\frac{U_i q_i+U_e q_e}{Q_w}.
\end{equation}
Combining this with Eq.~\eqref{eq: effective distance}, we can directly obtain the force exerted on the ion (electron) by the wire with potential $U_w^\prime$, namely, 
\begin{equation}\label{eq:F_i F_e}
    F_{i(e)}=\frac{q_{i(e)} U_w^\prime}{D_{i(e)}}.
\end{equation}
Substituting Eqs.~\eqref{eq:eq:U ion electron} and~\eqref{eq: U wire prime} into Eq.~\eqref{eq:F_i F_e} and ignoring constant forces, we arrive at 
\begin{gather}\label{eq:F_i final}
    F_i=-x_i \frac{q_i^2}{C_w D_i^2}-x_e\frac{q_iq_e}{C_w D_i D_e},\\
    \label{eq:F_e final}
    F_e=-x_i \frac{q_iq_e}{C_w D_i D_e} -x_e \frac{q_e^2}{C_w D_e^2}.
\end{gather}

We then take into account the trapping potentials of the ion and the electron and obtain
\begin{gather}
    m_i \Ddot{x}_i+m_i\omega_i^2x_i -F_i=0,\\
    m_e \Ddot{x}_e+m_e\omega_e^2[1+\eta\cos(\omega_d t)]x_e -F_e=0,
\end{gather}
where $m_{i(e)}$ and $\omega_{i(e)}$ represent the mass and harmonic trapping frequency of the ion (electron), respectively, whereas $\eta$ and $\omega_d$ denote the depth and frequency of the driving field applied to the trapped electron, respectively. Substituting $F_i$ and $F_e$ with the expressions given by Eqs.~\eqref{eq:F_i final} and~\eqref{eq:F_e final}, respectively, we have
\begin{gather}
    m_i \Ddot{x}_i+m_i\omega_i^2(1+\alpha_i)x_i+\gamma x_e=0,\\
    m_e \Ddot{x}_e+m_e\omega_e^2[1+\eta\cos(\omega_d t)+\alpha_e]x_e+\gamma x_i=0,
\end{gather}
where we have introduced the two shorthands
\begin{gather}
    \alpha_{i(e)}=\frac{q_{i(e)}^2}{C_w m_{i(e)} \omega_{i(e)}^2 D_{i(e)}^2},\label{eq:alphas}\\
    \gamma=\frac{q_i q_e}{C_wD_i D_e}.\label{eq:coupling constant}
\end{gather}
Therefore, the three parts of the total Hamiltonian introduced in Eq.~\eqref{eq:H} are explicitly given by 
\begin{gather}
    H_i=\frac{p_i^2}{2m_i}+\frac{1}{2}m_i\omega_i^2(1+\alpha_i) x_i^2,\\
    H_e=\frac{p_e^2}{2m_e}+\frac{1}{2}m_e\omega_e^2 x_e^2[1+\alpha_e+\eta\cos(\omega_d t)],\\
    H_I=\gamma x_i x_e, \label{eq:H_interaction}
\end{gather}
where $p_{i(e)}$ represents the momentum of the ion (electron).

\begin{figure}[tb]
    \centering
    \includegraphics{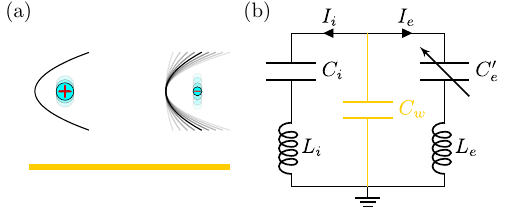}
    \caption{(a) Schematic of separately trapped ions and electrons connected via a conductor wire (golden line). The trapping potential of electron is properly modulated to ensure an efficient coupling. (b) Equivalent circuit of the system shown in (a). The ion motion is equivalent to the LC circuit with capacitance $C_i$ and inductance $L_i$, whereas the electron motion is equivalent to that with variable capacitance $C_e^\prime$ and inductance $L_e$. The conductor wire with capacitance $C_w$ to the ground serves as a bridge between the two LC circuits.}
    \label{fig1} 
\end{figure}

The equivalent circuit of this system is represented in \fig{fig1}(b), where the electric current $I_{i(e)}$ corresponds to the ion (electron) velocity~\cite{winelandPrinciplesStoredIon1975}, $L_{i(e)}=m_{i(e)} D_{i(e)}^2/q_{i(e)}^2$ and $C_{i(e)}=[\omega_{i(e)}^2 L_{i(e)}]^{-1}$ denote the equivalent inductance and capacitance of the ion (electron)~\cite{heinzenQuantumlimitedCoolingDetection1990,anCouplingTwoLasercooled2022}, respectively, and one has $1/C_e^\prime=1/C_e +1/C_d$, with the capacitance $C_d=[\eta\cos(\omega_d t)\omega_e^2 L_e]^{-1}$ accounting for the driving field.
Using the equivalent circuit elements, one obtains $\alpha_{i(e)}=C_{i(e)}/C_w$ and $\gamma=\pm\sqrt{m_i m_e/L_i L_e C_w^2}$. Here, we briefly note that $\omega_e/\omega_i\gg1$ is expected under typical experimental conditions. Furthermore, this model can be extended straightforwardly to the case of multiple particles in each trap by considering the center-of-mass motion and making the replacements $q_{i(e)}\to N_{i(e)} q_{i(e)}$ and $m_{i(e)}\to N_{i(e)} m_{i(e)}$, where $N_{i(e)}$ is the number of ions (electrons)~\cite{winelandPrinciplesStoredIon1975,heinzenQuantumlimitedCoolingDetection1990,bohmanSympatheticCoolingTrapped2021,willSympatheticCoolingSchemes2022}.

\subsection{Effective coherent coupling}
To illustrate the effective coherent coupling between the ions and electrons in such a  setup, we rewrite the Hamiltonian~\eqref{eq:H} using annihilation and creation operators.
Using the annihilation operator $a=\sqrt{m_i\omega_i'/2\hbar}(x_i+ip_i/m_i\omega_i^\prime)$ and suppressing the zero-point energy, the Hamiltonian of the ion motion can be directly rewritten as $H_i=\hbar\omega_i^\prime a^\dagger a$, with the effective ion frequency $\omega_i^\prime=\omega_i\sqrt{1+\alpha_i}$. 

The electron Hamiltonian, on the other hand, requires the transformation to an appropriate reference frame before it can be cast in the form $\tilde{H}_e=\hbar\partial_t\theta(t) b^\dagger b$,
if we also omit the zero-point energy and define $\theta(t)$ as the phase of a special solution
\begin{equation}
f(t)=r(t)\exp[i\theta(t)]
\end{equation}
of the Mathieu equation that constitutes the classical equation of motion under $H_e$~\cite{brownQuantumMotionPaul1991,glauber1992,leibfriedQuantumDynamicsSingle2003,kielpinskiQuantumInterfaceElectrical2012,kafriDynamicsIonCoupled2016}. 
The annihilation operator of the electron motion thus reads
\begin{equation}
    b=\sqrt{\frac{m_e W}{2\hbar}}\qty(x_e+\frac{i}{m_e W}p_e)
\end{equation}
and is time independent, since the Wronskian $W=(f^*\partial_t f-f\partial_t f^*)/2i=r^2\partial_t\theta$ satisfies $\partial_tW=0$~\cite{kafriDynamicsIonCoupled2016}.

We note that the position of the electron, $x_e$, is rescaled to $r x_e$ in this reference frame, leading to the coupling
\begin{equation}
    \tilde{H}_I=r\gamma x_i x_e=\frac{\hbar r\gamma}{2\sqrt{m_i m_e \omega_i^\prime W}}(a+a^\dagger)(b+b^\dagger).
\end{equation}
In the interaction picture with respect to $H_i$ and $\tilde{H}_e$, where the annihilation operators transform according to $a\to a \exp(-i\omega_i^\prime t)$ and $b\to b \exp[-i\theta(t)]$, we find the coupling Hamiltonian
\begin{equation}\label{eq: HI int}
    \tilde{H}^\mathrm{int}_I=\frac{\hbar \gamma}{2\sqrt{m_i m_e \omega_i^\prime W}}(a e^{-i\omega_i^\prime t}+a^\dagger e^{i\omega_i^\prime t})(b f^*+b^\dagger f).
\end{equation}

The general form of $f(t)$ can be obtained by solving the 
classical motion of the electron under $H_e$, which is governed by
\begin{equation}
    \ddot{x}_e + \omega_e ^{\prime 2} \qty[1+\eta^\prime\cos(\omega_d t)] x_e=0,
\end{equation}
where $\omega_e^\prime=\omega_e\sqrt{1+\alpha_e}$ is the effective electron frequency and $\eta^\prime=\eta/(1+\alpha_e)$ is the effective driving depth.
This equation of motion can be transformed into the standard Mathieu equation~\cite{NIST:DLMF}
\begin{equation}\label{eq:standard Mathieue}
    \frac{d^2x_e}{d\xi^2}+[A-2Q\cos(2\xi)]x_e=0,
\end{equation}
using the substitutions
\begin{equation}
    \xi=\frac{\omega_d t}{2},\quad A=\frac{4\omega_e^{\prime2}}{\omega_d^2},\quad Q= -\frac{A\eta^\prime}{2}.
\end{equation}
The Floquet solution of Eq.~\eqref{eq:standard Mathieue} is given by
\begin{equation}\label{eq:Floquet solution}
    f(\xi)=e^{i\mu\xi}\sum_{k=-\infty}^{\infty}{c_k e^{i2k\xi}},
\end{equation}
where $\mu$ is the characteristic exponent for the standard form of the Mathieu equation, and $c_k$ is its $k$th Fourier coefficient~\cite{NIST:DLMF,abramowitzHandbookMathematicalFunctions1965,shuObservationFloquetRaman2018} and we thereby find
\begin{equation}\label{eq:ft}
    f(t)=e^{i\mu \omega_d t /2}\sum_{k=-\infty}^\infty{c_k e^{ik\omega_d t}}.
\end{equation}
Then, under the resonance condition
\begin{equation}\label{eq:resonant condition}
    (\mu+2k)\omega_d/2=\omega_i^\prime,
\end{equation}
the $k$th Fourier component of the electron motion will be resonant with the ion motion.
We can therefore perform a rotating-wave approximation discarding highly oscillating terms in Eq.~\eqref{eq: HI int} and finally obtain
\begin{equation}\label{eq:H_int_RWA}
    \tilde{H}^\mathrm{int}_I\approx \hbar g_k(ab^\dagger+a^\dagger b),
\end{equation}
where the coupling strength has the form
\begin{equation}\label{eq:g_k}
    g_k=\frac{\gamma c_k}{2\sqrt{m_i m_e \omega_i^\prime W}},
\end{equation}
with $c_k/\sqrt{W}$ decided by the driving-field parameters. The Hamiltonian~\eqref{eq:H_int_RWA} describes a coherent excitation exchange between ion and electron \cite{brownCoupledQuantizedMechanical2011}, enabling the sympathetic or exchange cooling of the electron motion~\cite{heinzenQuantumlimitedCoolingDetection1990,willSympatheticCoolingSchemes2022}.

\begin{figure*}[tb]
    \centering
    \includegraphics{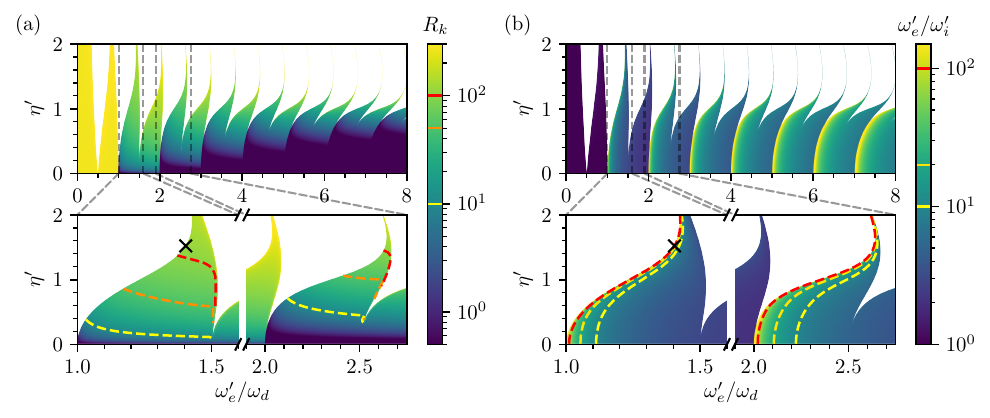}
    \caption{(a) The relative coupling strength $R_k$ for $k=0$ across various values of $\eta^\prime$ and $\omega_e^\prime/\omega_d$. 
        (b) The ratio between $\omega_e^\prime$ and $\omega_i^\prime$ resulted from the resonant condition for $k=0$ across various values of $\eta^\prime$ and $\omega_e^\prime/\omega_d$. 
        The areas with white color in both (a) and (b) correspond to the regions with non-real characteristic exponent $\mu$, where the electron motion is unstable.
        The enlarged panels in both (a) and (b) are depicted with contour lines in the regions bounded by $\mu=0$ and $\mu=1$.
        The values represented by the contour lines are indicated in the colorbars.
        The black cross denotes the typical parameters that we use in our further discussion.
    }
    \label{fig2} 
\end{figure*}

\begin{figure*}[tb]
    \centering
    \includegraphics{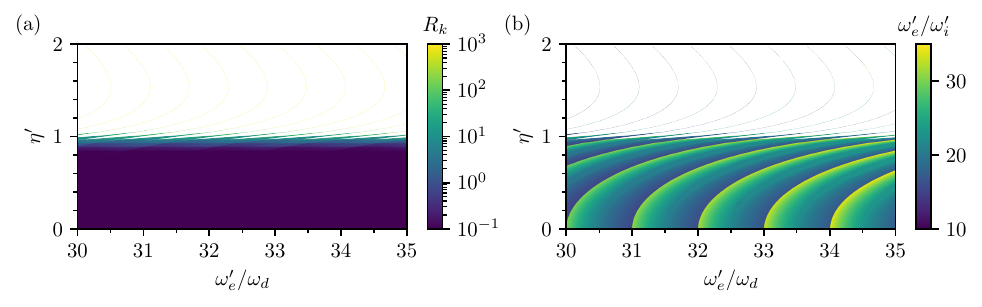}
    \caption{(a) The relative coupling strength $R_k$ with $k=1$ across various values of $\eta^\prime$ and $\omega_e^\prime/\omega_d$. 
        (b) The ratio of $\omega_e^\prime$ to $\omega_i^\prime$ resulted from the resonant condition with $k=1$ across various values of $\eta^\prime$ and $\omega_e^\prime/\omega_d$.}
    \label{fig:k=1}
\end{figure*}

\section{Driving-field parameters}
We now analyze the relation between the driving-field parameters and the coupling strength under the resonance condition~\eqref{eq:resonant condition}.
To illustrate the enhancement of the coupling strength with respect to the ion-wire-ion coupling, we define the relative coupling strength
\begin{equation}\label{eq:Rk}
    R_k=\abs{\frac{g_k}{g_{ii}}}=D_k\sqrt{\frac{m_i}{m_e}},
\end{equation}
where $g_{ii}=0.5\gamma/\omega_i^\prime m_i$ is the ion-wire-ion coupling strength, and $D_k=\abs{c_k}\sqrt{\omega_d(\mu+2k)/2W}$ is a factor depending on the effective driving depth $\eta^\prime$ and the ratio of the effective electron frequency $\omega_e^\prime$ to the driving frequency $\omega_d$.
Equation~\eqref{eq:Rk} shows that the enhancement is proportional to the square root of the mass ratio of ion to electron $\sqrt{m_i/m_e}$, which is, for example, around $269$ for $^{40}\mathrm{Ca}^+$ ions.

We then give a detailed derivation to show the dependence of $D_k$ to the driving-field parameters.
The characteristic exponent $\mu$ and the Fourier coefficients $c_k$ in Eq.~\eqref{eq:Floquet solution} can be numerically calculated for the specified parameters $A$ and $Q$ if we restrict $\Re(\mu)\in[0,2)$ and $\sum{\abs{c_k}^2}=1$~\cite{coissonMathieuFunctionsNumerical2009}.
With $\mu$ and $c_k$, the Wronskian with respect to $\xi$, $W_\xi=(f^*\partial_\xi f-f\partial_\xi f^*)/2i$, can be calculated by 
\begin{equation}
\begin{aligned}
    W_\xi&=\frac{1}{2i}(f^*\partial_\xi f-f\partial_\xi f^*)|_{\xi=0}\\
    &=\sum_{k,k^\prime}{c_k c_k^\prime(\mu+k+k^\prime)}.
\end{aligned}
\end{equation}
Then we can obtain the Wronskian with respect to $t$ by
\begin{equation}\label{eq:W}
    W=\frac{\omega_d}{2}W_\xi.
\end{equation}
The parameter $D_k$ in Eq.~\eqref{eq:Rk} can then be expressed with $W_\xi$ as 
\begin{equation}\label{eq:Dk}
    D_k=\abs{c_k}\sqrt{\frac{\mu+2k}{W_\xi}}.
\end{equation}
Therefore, for any given $\omega_e^\prime/\omega_d$ and $\eta^\prime$, we can first calculate the corresponding $A$ and $Q$ and then obtain $D_k$ using Eq.~\eqref{eq:Dk}.
Specifying $\sqrt{m_i/m_e}$ subsequently, the relative coupling strength $R_k$ can be obtained.

\figg{fig2}(a) illustrates the relative coupling strength $R_k$ for different values of $\eta^\prime$ and $\omega_e^\prime/\omega_d$ with $k=0$ and $m_i$ equal to the mass of $^{40}$Ca$^+$.
The colored (gray-scale) areas in \fig{fig2}(a), correspond to real values of the characteristic exponent $\mu$ and represent the electron-motion regions where stable solutions of the Mathieu equation exist, whereas the white areas correspond to non-real $\mu$ and denote the regions of unstable electron motion. In addition, the boundaries of the stability regions correspond to integer $\mu$~\cite{NIST:DLMF}.
As shown in \fig{fig2}(a), with increasing values of $\eta^\prime$, there is a concomitant increase in the relative coupling strength $R_k$, accompanied by a reduction in the size of the stable region. This puts a constraint on increasing $\eta^\prime$.
We further observe that, with increasing $\omega_e^\prime/\omega_d$, the relative coupling strength $R_k$ exhibits an overall decline across stable regions. This trend is illustrated by the expanded dark regions in \fig{fig2}(a) as $\omega_e^\prime/\omega_d$ increases. 
The enlarged panel of \fig{fig2}(a) compares two regions both delimited by a left boundary with $\mu=0$  and a right boundary with $\mu=1$.
The contour lines in the enlarged panel of \fig{fig2}(a) overall shift upwards for the region with larger $\omega_e^\prime/\omega_d$, indicating that a higher value of $\eta^\prime$ is required to achieve the same relative coupling strength $R_k$ when parameters fall within the region with larger $\omega_e^\prime/\omega_d$.

Using Eq.~\eqref{eq:resonant condition}, the ratio $\omega_e^\prime/\omega_i^\prime$ can be expressed as 
\begin{equation}\label{eq:omega_ei}
    \frac{\omega_e^\prime}{\omega_i^\prime}=\frac{\omega_e^\prime}{\omega_d}\frac{2}{\mu+2k}.
\end{equation}
Given that $\mu$ depends solely on $\eta^\prime$ and $\omega_e^\prime/\omega_d$, Eq.~\eqref{eq:omega_ei} can be used to determine the ratio $\omega_e^\prime/\omega_i^\prime$ for any given $k$, $\eta^\prime$ and $\omega_e^\prime/\omega_d$.
\figg{fig2}(b) presents the numerically calculated $\omega_e^\prime/\omega_i^\prime$ for $k=0$ across various values of $\eta^\prime$ and $\omega_e^\prime/\omega_d$.
For regions corresponding to conventional experimental configurations with $\omega_e^\prime/\omega_i^\prime\gg1$, it is found that $\mu \to 0$ and $\omega_e^\prime>\omega_d$.
Since $\omega_e^\prime/\omega_i^\prime$ is proportional to $\omega_e^\prime/\omega_d$, the light yellow areas in \fig{fig2}(b) expand as $\omega_e^\prime/\omega_d$ increases.
The enlarged panel of \fig{fig2}(b) clearly shows that the contour lines overall shift to the center of the stable region for those with larger $\omega_e^\prime/\omega_d$, indicating that a larger $\mu$ can be used to achieve the same $\omega_e^\prime/\omega_i^\prime$ when parameters fall within the region with larger $\omega_e^\prime/\omega_d$.

For $k\geq1$ and conventional experimental configurations, where $\omega_e^\prime/\omega_i^\prime\gg1$, Eq.~\eqref{eq:omega_ei} leads to  
\begin{equation}\label{eq:omega_eg_gg_k}
    \frac{\omega_e^\prime}{\omega_d}\gg k.
\end{equation}
\fig{fig:k=1} takes $k=1$ as an example, showing the relative coupling strength $R_k$ under such condition. In this case, $R_k$ only gains significant magnitude in the very narrow stable region where $\eta^\prime>1$, which may result unstable trapping in experiments.
Similar conclusion can also be found for the case of $k>1$.
Therefore, we mainly focus on the case of $k=0$.

The black cross in \fig{fig2} denotes a typical set of the potential parameters.
This typical set is calculated in a reverse way.
We first specify $\omega_d=25\omega_i^\prime$ and $k=0$, which leads to $\mu=2\omega_i^\prime/\omega_d=0.08$ according to the resonance condition of Eq.~\eqref{eq:resonant condition}.
We then set $Q=-6.0$ and calculate $A$ reversely from $\mu$ and $Q$ by constraining $A$ into the specific stable region.
In our case, we have $A\approx7.88$.
Then we can calculate $\eta^\prime=-2Q/A\approx1.52$ and $\omega_e^\prime=\omega_d\sqrt{A}/2\approx35\omega_i^\prime$.
With this set of parameters, the relative coupling strength $R_0$ between single $^{40}$Ca$^+$ ions and single electrons is approximately $111.5$ and the corresponding coupling strength may reach kilohertz regime for the Paul trap geometry demonstrated in Ref.~\cite{anCouplingTwoLasercooled2022}.

\section{Electron-mediated ion-ion coupling}
\begin{figure}[tb]
    \centering
    \includegraphics{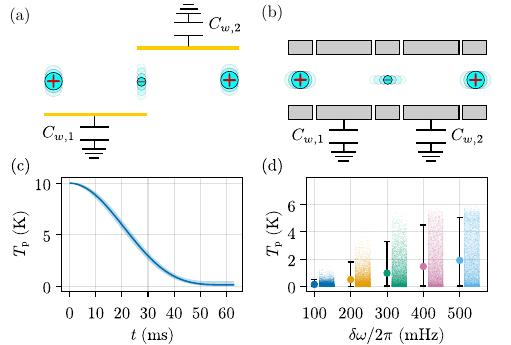}
    \caption{(a) Schematic of electron-mediated ion-ion coupling for Paul traps. (b) Schematic of electron-mediated Be-proton coupling for Penning traps. The gray blocks represent cross-sectional views of the ring electrodes. $C_{w,1(2)}$ in (a) and (b) is the capacitance of the common electrode 1{(2)} to ground. (c) Proton temperatures within one exchange ($t\in[0,t_\mathrm{ex}]$) for $\delta \omega=(2\pi)100$~mHz. The solid line represents the proton average temperature and the band represents the $5\%$-$95\%$ range of the proton temperatures.  (d) Proton temperatures at $t_\mathrm{ex}$ for $\delta \omega=(2\pi)[100,200,...,500]$~mHz. Large scatters, error bars and small scatters denote the proton average temperatures, the $5\%$-$95\%$ ranges of the proton temperatures and the proton temperature of every trajectory, respectively. }
    \label{fig3} 
\end{figure}

We now show that one can enhance the coupling between two separately trapped ions via an electron-mediated image-current interaction.
The schematics of the model for Paul and Penning traps are shown in \fig{fig3}(a) and (b), respectively. In both setups, two separately trapped ion clouds are connected to the central electrons via two common electrodes.
In Paul traps, the axial direction is perpendicular to the surfaces of the conductor wires [golden lines in \fig{fig3}(a)]. In Penning traps, the axial direction is horizontally along the ring electrodes [gray blocks in \fig{fig3}(b)].
This electron-mediated ion-ion coupling could be appealing for applications, such as qubit interconnection~\cite{brownCodesigningScalableQuantum2016,anCouplingTwoLasercooled2022} and sympathetic cooling~\cite{bohmanSympatheticCoolingProtons2018,bohmanSympatheticCoolingTrapped2021,willSympatheticCoolingSchemes2022}, and may also bring a revival of exchange cooling~\cite{heinzenQuantumlimitedCoolingDetection1990} due to the enhanced coupling rate.

The Hamiltonian of the ion-electron-ion system in the interaction picture can be written as
\begin{equation}
\label{eq:H iei}
    \tilde{H}^\mathrm{int}_{iei}=\sum_{j=1,2}{\hbar g_j ( e^{-i\delta_j t}a_j^\dagger b +h.c.)},
\end{equation}
where $g_j$ is the coupling strength between the center-of-mass motion of the electrons and the ions in the $j$th trap, as described by the annihilation operator $a_j$, and  $\delta_j$ represents the corresponding frequency detuning.

As for quantum information processing, we envisage that such ion-ion coupling mediated by electrons can be utilized to construct deterministic entanglement of separately trapped single ions. 
First, we initialize two laser coolable and separately trapped single ions into the motional state $\ket{1,0}$ at $t=0$.
Assuming that, in Eq.~\eqref{eq:H iei}, $g_1=g_2=g$ and 
\begin{equation}
    \delta_1=\delta_2=\delta=\pm g\sqrt{\frac{8n^2}{2n+1}},\quad n=1,\ 3,\ 5,\ ...,
\end{equation}
the motional energies of the ions are fully exchanged, regardless of the initial state of the electrons, after the swap time
\begin{equation}\label{eq: tau swap}
    \tau_\mathrm{swap}=\frac{\pi}{\abs{g}}\sqrt{\frac{2n+1}{2}}.
\end{equation}
At $\tau_\mathrm{swap}/2$, the motional quantum state of the two ions is in a maximally-entangled state, 
namely in one of the two N00N states $(|1,0\rangle\pm i|0,1\rangle)/\sqrt{2}$~\cite{dowlingQuantumOpticalMetrology2008a} (see Appendix~\ref{appendix: ion ion coupling}).
The detuning $\delta$ is intentionally added to make the electron motion return to its initial state at $\tau_\mathrm{swap}/2$, which is similar to the M{\o}lmer-S{\o}rensen gate for trapped ions \cite{daniilidisQuantumInformationProcessing2013,sorensenEntanglementQuantumComputation2000,roosIonTrapQuantum2008}.
For the ion-ion coupling with coupling strength $g_{ii}$, the swap time is $\pi/2g_{ii}$~\cite{brownCoupledQuantizedMechanical2011}.
Therefore, the gain of the ion-electron-ion coupling relative to the ion-ion coupling under similar experimental parameters can be expressed as
\begin{equation}\label{eq: coupling gain}
    \mathrm{gain}=\frac{\pi/2g_{ii}}{\tau_\mathrm{swap}} =R_0 \sqrt{\frac{1}{2(2n+1)}},
\end{equation}
where $R_0$ is given by Eq.~\eqref{eq:Rk} with Fourier order $k=0$.
For two $^{40}\mathrm{Ca}^+$ ions trapped separately, the coupling gain is approximately $45.5$ with $n=1$ and the driving parameters indicated by the black cross in \fig{fig2}.

For a near symmetric and resonant case, the two ion clouds in separated traps will almost fully exchange their energy at 
\begin{equation}
    t_\mathrm{ex}=\frac{\pi}{\sqrt{g_1^2+g_2^2}},
\end{equation}
regardless of the initial state of the electrons, which can be utilized to achieve exchange cooling.
To illustrate this, we numerically simulate the electron-mediated energy exchange process between $^9$Be$^+$ ions and a single proton in the classical regime.
All simulation parameters are referenced from well-developed Penning trap experiments~\cite{bohmanSympatheticCoolingTrapped2021,willSympatheticCoolingSchemes2022,willImagecurrentMediatedSympathetic2023}.
The effective electrode distances are set to $-D_\mathrm{1,Be}=D_{1,e}=-D_{2,e}=D_\mathrm{2,P}=3.2$~mm, where the first subscript denotes the electrode number, the second subscript denotes the particle species, and $\pm$ is decided by the relative position of the particles to the electrodes. 
The capacitances of the electrodes are set to $C_{w,1}=C_{w,2}=5.5$~pF.
The effective trapping frequencies of Be and proton are $\omega^\prime_\mathrm{Be}=\omega^\prime_\mathrm{P}=(2\pi)354.25$~kHz,
and the driving frequency, effective driving depth and effective trapping frequency of the electron trap, denoted by the black cross in \fig{fig2}, are $\omega_d=25\omega^\prime_\mathrm{Be}$, $\eta^\prime\approx 1.52$, and $\omega^\prime_e\approx35\omega^\prime_\mathrm{Be}$, respectively.
The ratio of the number of Be ions $N_\mathrm{Be}$ to the number of protons $N_\mathrm{P}$ is chosen to be approximately equal to the ratio of $m_\mathrm{Be}$ to $m_\mathrm{P}$, which insures that $g_1$ and $g_2$ in Eq.~\eqref{eq:H iei} are approximately equal in the symmetric trap configuration.
For the single-proton case, we have $N_\mathrm{Be}=9$.
The electron number $N_e$ is set in the simulation to $1000$, a value that is attainable in experimental settings~\cite{paascheInstabilitiesElectronCloud2003,satyajitLoadingDetectionNumber2009,dyavappaDependenceConfinementTime2017}.
The coupling constants $\gamma_{\mathrm{Be},e}$ and  $\gamma_{\mathrm{P},e}$ can be obtained with Eq.~\eqref{eq:coupling constant}.
Using Eq.~\eqref{eq:g_k}, we then obtain the corresponding coupling rates $g_1\approx g_2\approx (2\pi) 5.6$~Hz.
The coupling gain, relative to the Be-Proton wire-mediated coupling with $N_\mathrm{Be}=1000$ and $N_\mathrm{P}=1$, can also be calculated using Eq.~\eqref{eq: coupling gain} with $n=0$ and considering the mass and particle number modifications to $R_0$. 
In this case, the coupling gain is approximately $37.5$.

Our simulation takes into account the relative frequency uncertainties by randomly adding normal-distributed detunings with standard deviation $\delta \omega$ to the electron and proton trap for every simulation trajectory, see Appendix~\ref{appendix: numerical exchange cooling}.
Other heating noise is negligibly small for a cryogenic Penning trap.
The simulation results are obtained from 10\,000 trajectories for each $\delta\omega$ with the Be cloud initially cooled to $0.5$~mK, near the Doppler limit~\cite{willSympatheticCoolingSchemes2022}, and the temperatures of the electron cloud and the single proton are both equal to $10$~K.
The relative initial phases of the proton and the electrons motion are randomly chosen from $[0,2\pi]$, see Appendix~\ref{appendix: numerical exchange cooling}.
The proton average temperature for $\delta \omega=(2\pi)100$~mHz is depicted by the solid line in \fig{fig3}(c).
After a single exchange, at $t_\mathrm{ex}\approx 63$~ms, the proton average temperature reaches $133.4$~mK.
Figure~\ref{fig3}(d) shows the dependence of the cooling performance on the frequency uncertainty. 
Even for $\delta \omega=(2\pi)500$~mHz, the temperature of the proton decreases $81\%$ on average within a single exchange ($63$~ms), reaching $1.9$~K.

\section{Conclusion}
In this work, we presented a wire-mediated scheme for coherent ion-electron coupling with a two-orders-of-magnitude-enhanced strength, compared to known wire-mediated ion-ion coupling.
For mere single ions and electrons, our analysis shows that the ion-electron coupling strength may reach 
kilohertz regime in feasible experimental setups.
With demonstrated techniques for ion-trap noise reduction, such as $\mathrm{Ar}^+$ bombardment~\cite{hite100FoldReductionElectricField2012} and operation at cryogenic temperatures~\cite{brandlCryogenicSetupTrapped2016}, the electron-mediated image-current coupling between ions would be competitive with existing remote-coupling methods based on ion shuttling~\cite{pinoDemonstrationTrappedionQuantum2021,kielpinskiArchitectureLargescaleIontrap2002} or photon linking~\cite{olmschenkQuantumTeleportationDistant2009,stephensonHighRateHighFidelityEntanglement2020} regarding quantum information processing.
Further numerical simulations of electron-mediated exchange cooling of single protons in low-noise Penning traps~\cite{goodwinResolvedSidebandLaserCooling2016,borchertMeasurementUltralowHeating2019} highlight the potential application of our scheme for precision measurements. Our work thereby lays the groundwork for the development of various hybrid platforms with promising prospects for electron-mediated coupling of ions to a broad range of systems, from microscopic particles like antiprotons to macroscopic devices like LC circuits~\cite{kotlerHybridQuantumSystems2017}.

\begin{acknowledgments}
    This work is supported by the National Natural Science Foundation of China (12161141011 and 12174138), the National Key R$\&$D Program of China (2018YFA0306600). 
\end{acknowledgments}

\appendix
\section{Details on electron mediated ion-ion coupling}\label{appendix: ion ion coupling}
Here, we give a detailed analysis on ion-electron-ion motion coupling for deterministic entanglement generation and exchange cooling in a rotating frame where the Hamiltonian is time-independent.

As for deterministic entanglement generation, assuming $g=g_1=g_2$ and $\delta=\delta_1=\delta_2$, we have
\begin{equation}
    \tilde{H}^\mathrm{int^\prime}_{iei}=\hbar\delta b^\dagger b+ g[(a_1^\dagger+a_2^\dagger)b + \textrm{H.c.}].
\end{equation}
Generally, the initial state can be written as 
\begin{equation}
    \ket{\psi(t)}=\sum_{n_1,n_2,n_3}c_{n_1,n_2,n_3}(t)\ket{n_1,n_2,n_3},
\end{equation}
where $c_{n_1,n_2,n_3}(t)$ is the probability amplitude of the state $\ket{n_1,n_2,n_3}$ with $n_1$, $n_2$ and $n_3$ representing the phonon numbers for ion 1, ion 2 and electron, respectively.
Using 
\begin{equation}
    i\hbar\frac{\partial\ket{\psi}}{\partial t}=\tilde{H}^\mathrm{int^\prime}_{iei}\ket{\psi},
\end{equation}
we have
\begin{align}
    i\dot{c}_{n_1,n_2,n_3}&=\delta n_3 c_{n_1,n_2,n_3}\nonumber\\
    &\quad+g\sqrt{n_3+1} \sqrt{n_1}c_{n_1-1,n_2,n_3+1}\nonumber\\
    &\quad+g\sqrt{n_3+1} \sqrt{n_2}c_{n_1,n_2-1,n_3+1}\nonumber\\
    &\quad+g\sqrt{n_3} \sqrt{n_1+1}c_{n_1+1,n_2,n_3-1}\nonumber\\
    &\quad+g\sqrt{n_3} \sqrt{n_2+1}c_{n_1,n_2+1,n_3-1},\label{eq:coeff}  
\end{align}
where $\dot{[\ ]}$ stands for $\frac{\partial}{\partial t}$.
For a total phonon number equal to $1$, we can simplify Eq.~\eqref{eq:coeff} to
\begin{gather}
    i\dot{c}_{100}=g c_{001},\\
    i\dot{c}_{010}=g c_{001},\\
    i\dot{c}_{001}=\delta c_{001}+g c_{100}+g c_{010}.
\end{gather}
With initial conditions $c_{100}=1$, $c_{010}=0$, and $c_{001}=0$,
the solution for the probability amplitudes is
\begin{align}
    c_{100}(t)&=\frac{1}{2}+\frac{e^{-\frac{1}{2}igt\delta_r}}{2}\cos{\left(\frac{1}{2}gt\sqrt{8+\delta_r^2}\right)}\nonumber\\
    &\quad+\frac{i\delta_r e^{-\frac{1}{2}igt\delta_r}}{2\sqrt{8+\delta_r^2}}\sin{\left(\frac{1}{2}gt\sqrt{8+\delta_r^2}\right)},
\label{eq:c100}\\
    c_{010}(t)&=-\frac{1}{2}+\frac{e^{-\frac{1}{2}igt\delta_r}}{2}\cos{\left(\frac{1}{2}gt\sqrt{8+\delta_r^2}\right)}\nonumber\\
    &\quad+\frac{i\delta_r e^{-\frac{1}{2}igt\delta_r}}{2\sqrt{8+\delta_r^2}}\sin{\left(\frac{1}{2}gt\sqrt{8+\delta_r^2}\right)},
\label{eq:c010}\\
    c_{001}(t)&=-\frac{2 i e^{-\frac{1}{2} i g t \delta_r  }}{\sqrt{8+\delta_r ^2}}\sin \left(\frac{1}{2}g t\sqrt{8+\delta_r ^2}\right),
\end{align}
where we defined $\delta_r=\delta/g$.
At the end of the evolution, $t=\tau$, we expect $c_{001}(\tau)=0$, which is fulfilled if
\begin{equation}\label{eq:mpi}
    \frac{1}{2}g \tau \sqrt{8+\delta_r^2}=m\pi, 
\end{equation}
with an integer $m$.
Substituting Eq.~\eqref{eq:mpi} into Eq.~\eqref{eq:c100} and Eq.~\eqref{eq:c010}, we thereby find
\begin{gather}
    c_{100}(\tau)=\frac{1}{2}\qty[1+(-1)^m e^{-\frac{1}{2}ig\tau\delta_r}],\label{eq:c100m}\\
    c_{010}(\tau)=\frac{1}{2}\qty[-1+(-1)^m e^{-\frac{1}{2}ig\tau\delta_r}]\label{eq:c010m}
\end{gather}
and if one now requires 
\begin{equation}\label{eq:npi/2}
    -\frac{1}{2}g\tau\delta_r=n\frac{\pi}{2},
\end{equation}
with another integer $n$ and combines this condition with Eq.~\eqref{eq:mpi}, one arrives at
\begin{gather}
    \delta=\pm g\sqrt{\frac{8n^2}{(2m)^2-n^2}},\label{eq:delta1}\\
    \tau=\frac{\pi}{2\abs{g}}\sqrt{\frac{(2m)^2-n^2}{2}},\label{eq:tau}
\end{gather}
where $m,n\in \mathbb{Z}$ and $\abs{2m}>\abs{n}$.

If $n$ is odd, we substitute Eq.~\eqref{eq:npi/2} into Eq.~\eqref{eq:c100m} and Eq.~\eqref{eq:c010m}, obtaining
\begin{gather}
    c_{100}(\tau)=\frac{1}{2}\qty{1+i\qty(-1)^{\qty[m+\frac{n-1}{2}]}},\\
    c_{010}(\tau)=\frac{1}{2}\qty{-1+i\qty(-1)^{\qty[m+\frac{n-1}{2}]}}
\end{gather}
which means the motion of the two ions is in one of the NOON states $\qty(\ket{1,0}\pm i\ket{0,1})/\sqrt{2}$ at $t=\tau$.
Without loss of generality, we can assume $n,m>0$. If we substitute $2m$ with $n+1$, we can obtain
\begin{gather}
    \delta=\pm g\sqrt{\frac{8n^2}{2n+1}},\label{eq:delta2}\\
    \tau=\frac{\pi}{2\abs{g}}\sqrt{\frac{2n+1}{2}},
\end{gather}
where we find that $\tau=\tau_\mathrm{swap}/2$ with $\tau_\mathrm{swap}$ defined by the Eq.~\eqref{eq: tau swap}.
On the other hand, if $n$ is even, we substitute Eq.~\eqref{eq:npi/2} into Eq.~\eqref{eq:c100m} and Eq.~\eqref{eq:c010m}, obtaining
\begin{gather}
    c_{100}(\tau)=\frac{1}{2}\qty{1+(-1)^{m+\frac{n}{2}}},\\
    c_{010}(\tau)=\frac{1}{2}\qty{-1+(-1)^{m+\frac{n}{2}}},
\end{gather}
which means that, at $t=\tau$, the two ions are either in their initial states for even $m+n/2$ or fully exchange their energies for odd $m+n/2$.
For the full-exchange case with even $m$ (namely $m+n/2$ is odd and $m$ is even), by rewriting Eq.~\eqref{eq:tau} as 
\begin{equation}\label{eq:tau_n_even}
    \tau=\frac{\pi}{\abs{g}}\sqrt{\frac{m^2-\qty(\frac{n}{2})^2}{2}},
\end{equation}
we find that $\tau/2$ in such a case is equivalent to $\tau$ with odd $n$. This means that the two ions not only exchange their energies at $t=\tau$ in such a case, but are also in a NOON state at $t=\tau/2$.
However, for the full-exchange case with odd $m$, $\tau/2$ is not equivalent to $\tau$ with odd $n$ and the non-integer $m/2$ results $c_\mathrm{001}(\tau/2)\neq0$, which means that the two ions are not in a NOON state at $\tau/2$.

In summary, we can substitute $n/2$ in Eq.~\eqref{eq:tau_n_even} with integer $n$ and obtain
\begin{equation}
    \tau=\tau_\mathrm{swap}=\frac{\pi}{\abs{g}}\sqrt{\frac{m^2-n^2}{2}},
\end{equation}
where $m,n\in \mathbb{Z}$, $\abs{m}>\abs{n}$ and $m+n$ is odd.
Only when $n$ is odd, the two ions are in a NOON states at $t=\tau_\mathrm{swap}/2$.
Setting $m=n+1$, we have 
\begin{equation}
    \tau_\mathrm{swap}=\frac{\pi}{\abs{g}}\sqrt{\frac{2n+1}{2}},
\end{equation}
which is Eq.~\eqref{eq: tau swap}.

Although we assume the initial phonon number of the electron is zero, the results we obtained here are actually not dependent on the initial phonon number of the electron.
We then discuss this argument in Heisenberg picture.
In Heisenberg picture, we have
\begin{align}
    a_1(t)&=\frac{1}{2}\qty[a_1(0)-a_2(0)+h(t)],\\
    a_2(t)&=\frac{1}{2}\qty[a_2(0)-a_1(0)+h(t)],\\
    b(t)&=e^{-igt\delta_r/2}b(0)\cos\left(\frac{gt\sqrt{8+\delta_r^2}}{2}\right)\\
    &\quad -\frac{i\qty[2a_1(0)+2a_2(0)+\delta_r b(0)]}{e^{igt\delta_r/2}\sqrt{8+\delta_r^2}}\sin\left(\frac{gt\sqrt{8+\delta_r^2}}{2}\right),\nonumber
\end{align} 
where
\begin{align}
    h(t)&=e^{-igt\delta_r/2}\qty[a_1(0)+a_2(0)]\cos\left(\frac{gt\sqrt{8+\delta_r^2}}{2}\right)\\
    &\quad +\frac{i\qty[\delta_r \qty(a_1(0)+a_2(0))-4b(0)]}{e^{igt\delta_r/2}\sqrt{8+\delta_r^2}}\sin\left(\frac{gt\sqrt{8+\delta_r^2}}{2}\right).\nonumber
\end{align}
At $t=\tau$, substituting $g\tau\sqrt{8+\delta_r^2}/2$ with Eq.~\eqref{eq:mpi}, we have
\begin{align}
    a_1(\tau)&=\frac{1}{2}a_1(0)\qty[1+(-1)^m e^{-ig\tau\delta_r/2}]\nonumber\\
    &\quad +\frac{1}{2}a_2(0)\qty[-1+(-1)^m e^{-ig\tau\delta_r/2}],\\
    a_2(\tau)&=\frac{1}{2}a_2(0)\qty[1+(-1)^m e^{-ig\tau\delta_r/2}]\nonumber\\
    &\quad +\frac{1}{2}a_1(0)\qty[-1+(-1)^m e^{-ig\tau\delta_r/2}],\\
    b(\tau)&=(-1)^m b(0)e^{-ig\tau\delta_r/2}.
\end{align} 
We see that $a_1(\tau)$ and $a_2(\tau)$ is not dependent on $b(0)$, which means that the results we mentioned before are not dependent on the initial motional state of the electron.

We then discuss the resonant case for exchange cooling.
For $g_1\neq g_2$ but $\delta_1=\delta_2=0$, defining $g=\sqrt{g_1^2+g_2^2}$,
in Heisenberg picture, we have
\begin{align}
    a_1(t)&=a_1(0)\qty[\frac{g_1^2}{g^2}\cos(gt)+\frac{g_2^2}{g^2}]+a_2(0)\frac{g_1 g_2}{g^2} [\cos(gt)-1]\nonumber\\
    &\quad -ib(0)\frac{g_1}{g}\sin(gt),\\
    a_2(t)&=a_2(0)\qty[\frac{g_2^2}{g^2}\cos(gt)+\frac{g_1^2}{g^2}]+a_1(0)\frac{g_1 g_2}{g^2} \qty[\cos(gt)-1]\nonumber\\
    &\quad-ib(0)\frac{g_2}{g}\sin(gt),\\
    b(t)&=b(0)\cos(gt)-i\frac{a_1(0)g_1+a_2(0)g_2}{g}\sin(gt).
\end{align}
These equations can be utilized to obtain the energy exchange of the classical motion by assuming the initial state to be a direct product state of coherent states, $\ket{\alpha_1,\alpha_2,\beta}$.
The center-of-mass energies of the separately trapped particles can be written as 
\begin{subequations}
\label{eq:E exchange}
    \begin{align}
        E_1(t)&=\hbar\omega\expval{a_1^\dagger(t) a_1(t)}=\hbar\omega\abs{\alpha_1(t)}^2,\\
        E_2(t)&=\hbar\omega\expval{a_2^\dagger(t) a_2(t)}=\hbar\omega\abs{\alpha_2(t)}^2,\\
        E_b(t)&=\hbar\omega_b\expval{b^\dagger(t) b(t)}=\hbar\omega_b\abs{\beta(t)}^2,\label{eq:E exchange electron}
    \end{align}
\end{subequations}
where
\begin{align}
        \alpha_1(t)&=\alpha_2\frac{g_1 g_2}{g^2} [\cos(gt)-1]+\alpha_1\frac{g_1^2}{g^2}\qty[\cos(gt)+\frac{g_2^2}{g_1^2}]\nonumber\\
        &\quad -i\beta\frac{g_1}{g}\sin(gt),\\
        \alpha_2(t)&=\alpha_1\frac{g_1 g_2}{g^2} [\cos(gt)-1]+\alpha_2\frac{g_2^2}{g^2}\qty[\cos(gt)+\frac{g_1^2}{g_2^2}]\nonumber\\
        &\quad -i\beta\frac{g_2}{g}\sin(gt),\\
    \beta(t)&=\beta\cos(gt)-i\frac{\alpha_1 g_1+\alpha_2 g_2}{g}\sin(gt),
\end{align}
$\omega$ is the frequency of the two separately trapped ions, and $\omega_b$ is the effective frequency of the electron under the driving field, which can be determined by the ratio of the electron average motional energy to the corresponding Fock number.
From Eqs.~\eqref{eq:E exchange}, we can find that, if $g_1=g_2$, at time $t_\mathrm{ex}=\pi/g$, the energies of ions are fully exchanged.

\begin{figure*}[tb]
    \centering
    \includegraphics{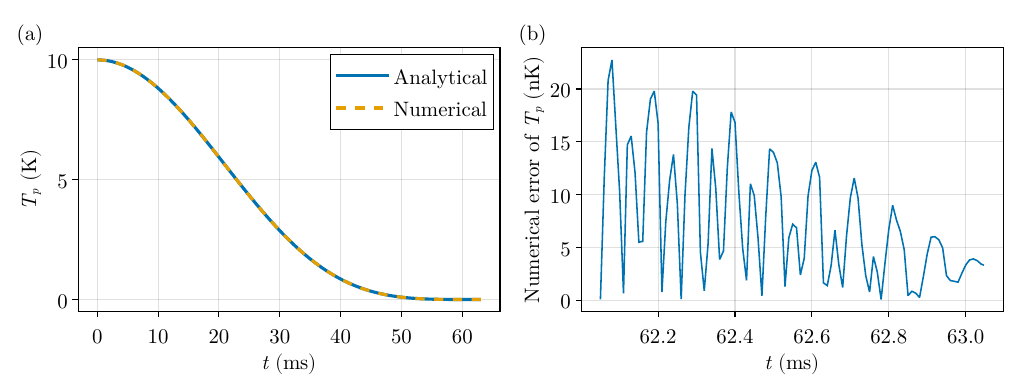}
    \caption{(a) Analytical (solid blue line) and numerical (dashed yellow line) solution of the proton temperature $T_p$ within a single exchange. 
             (b) Numerical error of the proton temperature around the end of the exchange.
            }
    \label{fig:exchange_convergence}
\end{figure*}

\section{Numerical details on exchange cooling}\label{appendix: numerical exchange cooling}
In the simulation of exchange cooling, we consider 9 $^9$Be$^+$ ions, 1000 electrons and a single proton.
The couplings between traps are mediated via common electrodes.
The classical motion of the system is governed by 
\begin{align}
    &\Ddot{x}_\mathrm{Be}+ \omega_\mathrm{Be}^{\prime 2} x_\mathrm{Be}+\frac{\gamma_{\mathrm{Be},e}}{N_\mathrm{Be}m_\mathrm{Be}} x_e\nonumber\\
    &=\Ddot{x}_e + \omega_e ^{\prime 2} [1+\eta^\prime\cos(\omega_d t)] x_e + \frac{\gamma_{\mathrm{Be},e}}{N_e m_e } x_\mathrm{Be}+\frac{\gamma_{\mathrm{P},e}}{N_e m_e } x_\mathrm{P}\nonumber\\
    &=\Ddot{x}_\mathrm{P}+ \omega_\mathrm{P}^{\prime 2} x_\mathrm{P}+\frac{\gamma_{\mathrm{P},e}}{N_\mathrm{P} m_\mathrm{P}} x_e=0,\label{eq:classical motion}
\end{align}
where $x_\mathrm{Be}$, $x_e$, $x_\mathrm{P}$ are the center-of-mass positions of the $^9$Be$^+$ ions, the electrons and the single proton, respectively.
The numerical integration of Eq.~\eqref{eq:classical motion} uses the 12th-order explicit adaptive Runge-Kutta-Nystr\"{o}m method provided by \verb|DifferentialEquations.jl|~\cite{rackauckasDifferentialEquationsJlPerformant2017a} with step-size control options $abstol=1.0\times 10^{-12}$, $reltol=1.0\times10^{-12}$, and number type \verb|Double64| provided by \verb|DoubleFloats.jl|~\cite{sarnoffDoubleFloats2022}. The number type \verb|Double64| has a significand of 106 bits, two times larger than that of the common number type \verb|Float64|.

\begin{figure}[tb]
    \raggedright
    \includegraphics{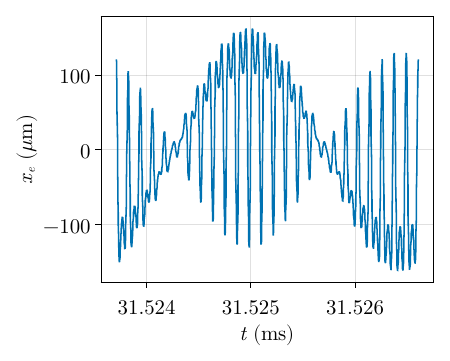}
    \caption{Electron dynamics around the midpoint of the exchange process.}
    \label{fig:exchange_electron}
\end{figure}

To make sure that the finite numerical precision does not negatively affect our simulation results, we compare our simulation results with the theoretical analysis.
Substituting the subscript $1$ and $2$ of $E$ and $\alpha$ in Eqs.~\eqref{eq:E exchange} with $\mathrm{Be}$ and $\mathrm{P}$ and setting $\beta=0$ and  $\alpha_\mathrm{Be}$ and $\alpha_\mathrm{P}$ in the same phase, we have
\begin{align}
    E_\mathrm{Be}(t)=\bigg|&\sqrt{E_\mathrm{P}(0)}\frac{g_1 g_2}{g^2} [\cos(gt)-1]\nonumber\\
    &+\sqrt{E_\mathrm{Be}(0)}\frac{g_1^2}{g^2}\qty[\cos(gt)+\frac{g_2^2}{g_1^2}]\bigg|^2,\\
    E_\mathrm{P}(t)=\bigg|&\sqrt{E_\mathrm{Be}(0)}\frac{g_1 g_2}{g^2} [\cos(gt)-1]\nonumber\\
    &+\sqrt{E_\mathrm{P}(0)}\frac{g_2^2}{g^2}\qty[\cos(gt)+\frac{g_1^2}{g_2^2}]\bigg|^2,\label{eq:E2 in phase}
\end{align}
which corresponds to the initial condition: $x_e(0)=x_\mathrm{Be}(0)=x_\mathrm{P}(0)=0$~$\mu$m, $\dot{x}_e(0)=0$, $\dot{x}_\mathrm{Be}(0)=\sqrt{2E_\mathrm{Be}(0)/m_\mathrm{Be}}$, and $\dot{x}_\mathrm{P}(0)=\sqrt{2E_\mathrm{P}(0)/m_\mathrm{P}}$.
Defining $T_\mathrm{Be}=E_\mathrm{Be}/k_B$ and $T_\mathrm{P}=E_\mathrm{P}/k_B$, we can obtain the solid blue line in \fig{fig:exchange_convergence}(a) using Eq.~\eqref{eq:E2 in phase} with $T_\mathrm{Be}(0)=0.5$~mK and $T_\mathrm{P}(0)=10$~K.
The dashed yellow line in \fig{fig:exchange_convergence}(a) shows the numerical result of $T_\mathrm{P}(t)$ within a single exchange, which is nearly identical to the theoretical one.
Around the end of the exchange, the energy deviation of the simulation result from the theoretical one is only a few nK$\times k_B$, as shown in \fig{fig:exchange_convergence}(b).
This numerical accuracy is sufficient for our case.
\figg{fig:exchange_electron} shows the center-of-mass position trajectory of the electrons around $t_\mathrm{ex}/2$, where the electrons reach the max amplitude during the exchange process.

Figure~\ref{fig3}(c) and (d) are obtained from 10\,000 trajectories for each standard deviation $\delta\omega$.
Besides the random detunings, every trajectory also takes account the random phases of the initial modes.
Other sources of noise are negligible in the cryogenic Penning trap~\cite{borchertMeasurementUltralowHeating2019,willSympatheticCoolingSchemes2022}.
Equations~\eqref{eq:E exchange} indicate that only relative initial phases will influence the energy of motion during exchange.
Therefore, we randomly add relative initial phases $\phi_\mathrm{P}$ and $\phi_e$ to the proton and electron initial state, respectively, and set the Be ion initial phase to $0$. 
For the single proton with initial temperature $T_\mathrm{P}$ and relative phase $\phi_\mathrm{P}$, we have 
\begin{gather}\label{eq:P position init}
    x_\mathrm{P}(0)=\frac{2k_B T_\mathrm{P}}{m_\mathrm{P}\omega_\mathrm{P}^{\prime2}}\sin(\phi_\mathrm{P}), \\
    \dot{x}_\mathrm{P}(0)=\frac{2k_B T_\mathrm{P}}{m_\mathrm{P}}\cos(\phi_\mathrm{P}).\label{eq:P velocity init}
\end{gather}
As for electrons, it is appropriate to define the electron temperature by the kinetic energy averaged over the infinite time $t^\prime$:
\begin{equation}\label{eq:e kin}
    E_{\mathrm{kin},e}=\lim_{t^\prime\to\infty}\frac{1}{t^\prime}\int_0^{t^\prime}{\frac{1}{2}m_e \dot{x}_e^2} \,dt=\frac{1}{2}k_B T_e
\end{equation}
To see the relationship between $T_e$ and the electron motion amplitude,
we rewrite Eq.~\eqref{eq:ft} into a form with sine functions:
\begin{equation}\label{eq:electron sin mathieu}
    x_e(t)=A_ef_s(t)=A_e\sum_{k=-\infty}^\infty{c_k\sin\qty[(\mu+2k)\frac{\omega_d t}{2}]},
\end{equation}
where $A_e$ is the amplitude of the electron motion.
Immediately, we can write down the electron velocity as
\begin{equation}
    \dot{x}_e(t)=\frac{A_e\omega_d}{2}\sum_{k=-\infty}^\infty{(\mu+2k)c_k\cos\qty[(\mu+2k)\frac{\omega_d t}{2}]}.
\end{equation}
With Parseval's theorem~\cite{kaplanAdvancedCalculus1991}, we obtain
\begin{equation}\label{Eq:coef square}
    \lim_{t^\prime\to\infty}\frac{1}{t^\prime}\int_0^{t^\prime}{\frac{1}{2}m_e \dot{x}_e^2} \,dt=
    \frac{A_e^2\omega_d^2}{8}\sum_{k=-\infty}^\infty{\abs{(\mu+2k)c_k}}^2.
\end{equation}
Substituting Eq.~\eqref{Eq:coef square} into Eq.~\eqref{eq:e kin} and using the definition $S=\sum_{k=-\infty}^\infty{\abs{(\mu+2k)c_k}}^2$, we obtain
\begin{equation}\label{eq:e_amp}
    A_e=\sqrt{\frac{8k_B T_e}{m_e\omega_d^2 S}}.
\end{equation}
With the relative phase $\phi_e$, we have
\begin{gather}\label{eq:e position init}
        x_e(0)=A_e\sin(\phi_e)\sum_{k=-\infty}^\infty{c_k},\\
        \dot{x}_e(0)=\frac{A_e\omega_d}{2}\cos(\phi_e)\sum_{k=-\infty}^\infty{(\mu+2k)c_k}.\label{eq:e velocity init}
\end{gather}

To sum up, for every simulation trajectory, we first randomly generate $\phi_\mathrm{P}$ and $\phi_e$ in the $[0,2\pi]$ range to obtain the initial motional state using Eqs.~\eqref{eq:P position init} and~\eqref{eq:P velocity init} and Eqs.~\eqref{eq:e position init} and~\eqref{eq:e velocity init}.
Then, we modify $\omega_e^\prime$ and $\omega_\mathrm{P}^\prime$ with two randomly generated relative detunings, whose standard deviations are both set to $\delta\omega$.
Next, we integrate Eqs.~\eqref{eq:classical motion} with numerical configurations mentioned above to obtain the original data for \fig{fig3}(b) and (c).

With Eq.~\eqref{eq:e_amp}, we can also calculate the effective frequency $\omega_b$ mentioned in Eq.~\eqref{eq:E exchange electron}.
Equation~\eqref{eq:e_amp} can be rewritten as
\begin{equation}\label{eq:E temperature amplitude}
    k_B T_e=\frac{A_e^2 m_e \omega_d^2 S}{8}.
\end{equation}
For a coherent state $\ket{\beta}$, we have
\begin{equation}
    k_B T_e=\hbar \omega_b \abs{\beta}^2.
\end{equation}
Therefore, we obtain
\begin{equation}\label{eq:E beta A_e}
    \hbar \omega_b \abs{\beta}^2=\frac{A_e^2 m_e \omega_d^2 S}{8}.
\end{equation}
According to the transformations made to obtain $\tilde{H}_e$, the classical amplitude $A_e$ is connected to the coherent state $\ket{\beta}$ by:
\begin{equation}\label{eq:A_e beta}
    A_e=\abs{\beta}\sqrt{\frac{2\hbar}{m_e W}},
\end{equation}
where $W$ is the Wronskian defined in Eq.~\eqref{eq:W}.
Substituting Eq.~\eqref{eq:A_e beta} into Eq.~\eqref{eq:E beta A_e} and utilizing Eq.~\eqref{eq:W}, we obtain 
\begin{equation}
    \omega_b =\frac{\omega_d S}{2W_\xi}.
\end{equation}
For the driving-field parameters that we choose in our simulation, $\omega_b$ is approximately $(2\pi)142$~MHz.

%

\end{document}